\documentclass[aps,prd,nofootinbib, two column]{revtex4-2}
\usepackage{graphicx}     
\usepackage{subcaption}   
\usepackage{caption}      
\usepackage{float}        

\usepackage[normalem]{ulem}
\usepackage{amsmath}
\usepackage{ulem}
\usepackage{graphicx}
\usepackage{epsfig}
\usepackage{bm}
\usepackage{enumitem} 
\usepackage{natbib}
\usepackage{pgfplots,mathtools}
\usepackage{hyperref}
\usepackage{amsmath}
\usepackage{braket}\usepackage{slashed}
\usepackage[compat=1.0.0]{tikz-feynman}
\usepackage{physics}
\usepackage{caption}
\usepackage{subcaption}
\usepackage{xfrac}
\usepackage{algorithm}

\newcommand{\bef}{\begin{figure}}
\newcommand{\eef}{\end{figure}}
\newcommand{\bc}{\begin{center}}
\newcommand{\ec}{\end{center}}

\newcommand{\be}{\begin{equation}}
\newcommand{\ee}{\end{equation}}
\newcommand{\bea}{\begin{eqnarray}}
\newcommand{\eea}{\end{eqnarray}}

\def\ba{\begin{eqnarray}}
\def\ea{\end{eqnarray}}

\newcommand{\orcid}[1]{\href{https://orcid.org/#1}{\includegraphics[width=8pt]
{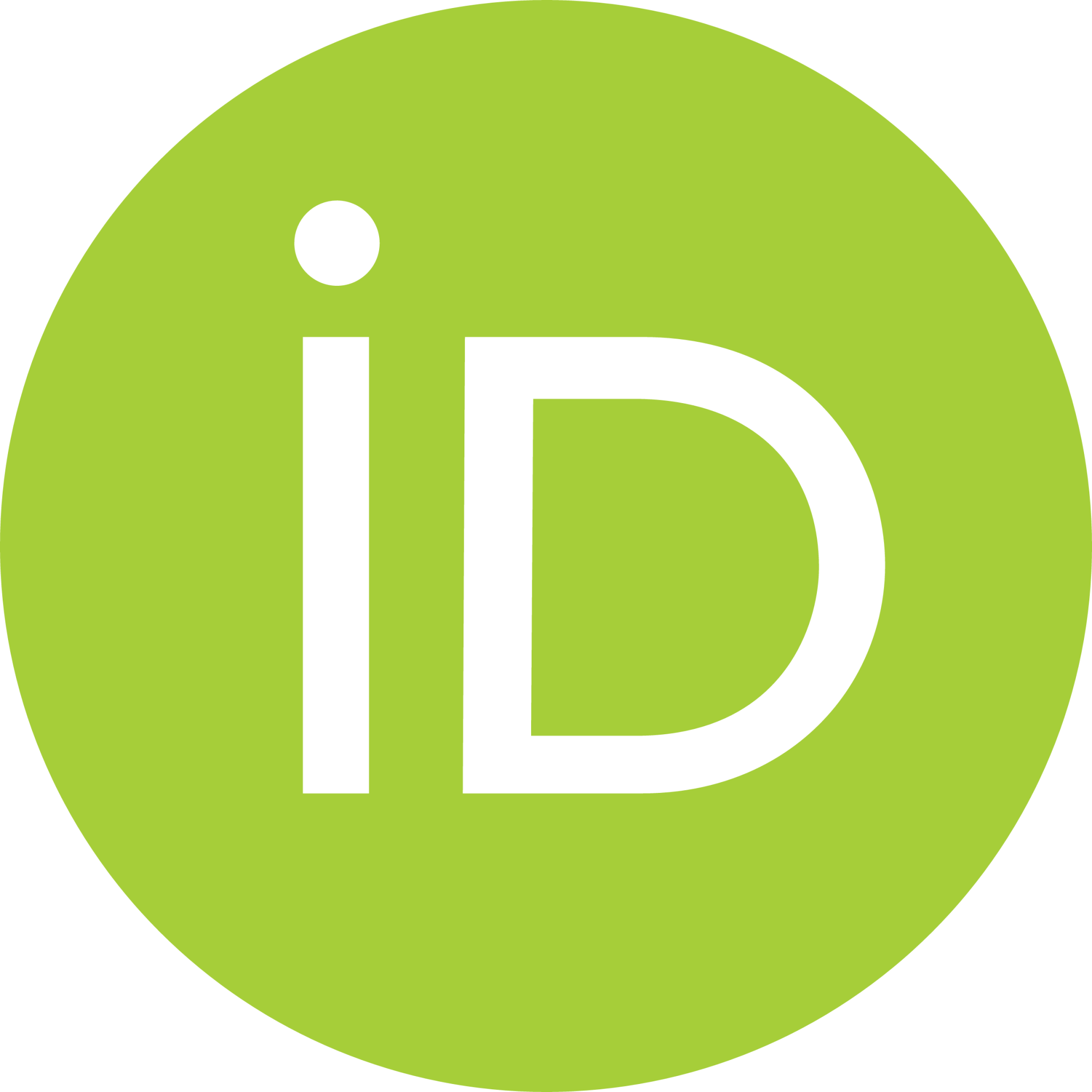}}}

\begin{document}
\title{Directed Flow of $\text{D}$ and $\text{B}$ Mesons in an Electrically and Chirally Conductive QGP at LHC Energies}

\author{
Ankit Kumar Panda\orcid{0000-0002-9394-6094}$^{1,2}$,
Pooja\orcid{0000-0002-6871-3475}$^{3,4}$,
Maria Lucia Sambataro\orcid{0009-0007-9018-661X}$^{5,6}$
Salvatore Plumari\orcid{0000-0002-3101-8196}$^{5,6}$, and
Santosh K. Das\orcid{0000-0003-3867-3158}$^{2}$
}

\affiliation{$^{1}$Key Laboratory of Quark and Lepton Physics (MOE) \& Institute of Particle Physics, Central China Normal University, Wuhan 430079, China}
\affiliation{$^{2}$School of Physical Sciences, Indian Institute of Technology Goa, Ponda 403401, Goa, India}

\affiliation{$^{3}$Department of Physics, University of Jyväskylä,  P.O. Box 35, 40014 University of Jyväskylä, Finland}
\affiliation{$^{4}$Helsinki Institute of Physics, P.O. Box 64, 00014 University of Helsinki, Finland}

\affiliation{$^{5}$Department of Physics and Astronomy “E. Majorana”, University of Catania, Via S. Sofia 64, I-95123 Catania, Italy}
\affiliation{$^{6}$Laboratori Nazionali del Sud, INFN-LNS, Via S. Sofia 62, I-95123 Catania, Italy}

\begin{abstract}
We investigate the directed flow of $\text{D}$ and $\text{B}$ mesons in the presence of electromagnetic fields incorporating finite electrical and chiral conductivities at LHC energies. The momentum evolution of heavy quarks in the quark–gluon plasma (QGP) is studied using Langevin dynamics, with their interactions with the medium described within the extended quasiparticle model ($QPM_p$) framework. The electromagnetic fields are obtained from analytical solutions of Maxwell’s equations that account for both electrical and chiral conductivities. These conductivities modify the space–time evolution of the electromagnetic fields, thereby influencing the splitting of the directed flow between mesons and anti-mesons.
However, the influence of chiral conductivity remains secondary to that of the electrical conductivity, and its impact on the directed flow is marginal in the current formulation. The results reveal that heavy mesons containing a charm quark develop a directed flow with a sign opposite to that of heavy mesons containing a bottom quark, with the magnitude being smaller for the latter. The present study indicates that a simultaneous experimental measurement of $v_{1}$ for heavy mesons containing both charm and bottom quarks would provide valuable insight into the electromagnetic-field origin of $v_{1}$ for the heavy quarks.

\end{abstract}
    
\maketitle
\section{Introduction}
High-energy heavy-ion collisions have been extensively studied over the past few decades to explore the fundamental constituents of matter—quarks and gluons. Such collisions create an extremely hot and dense medium of deconfined quarks and gluons, known as the quark–gluon plasma (QGP)~\cite{Bjorken:1982qr,BRAHMS:2004adc,STAR:2005gfr,PHENIX:2004vcz}. This exotic state of matter is believed to have existed a few microseconds after the Big Bang, making heavy-ion collisions a unique laboratory for investigating the strong interaction under extreme conditions.
Heavy quarks~\cite{Rapp:2018qla,Das:2024vac,He:2022ywp,Dong:2019unq,Aarts:2016hap,Cao:2018ews, Khowal:2021zoo, Ruggieri:2022kxv, Pooja:2022ojj, Pooja:2023gqt, Pooja:2024rnn}—primarily charm and bottom—serve as excellent probes of the QGP. Due to their large masses, they are produced predominantly during the earliest stages of the collision and experience relatively long thermalization times, allowing them to retain information about the medium through which they traverse. Moreover, since their masses are much greater than the typical temperature of the medium ($M \gg T$), their thermal production is exponentially suppressed by the Boltzmann factor ($\sim e^{-M/T}$), leading to approximate flavor conservation throughout the QGP evolution. As a result, heavy quarks encode valuable information about both the initial conditions and the subsequent transport properties of the medium.
A central objective of heavy-quark studies is the determination of the spatial diffusion coefficient, $D_s$, which quantifies their coupling strength to the bulk medium. Experimentally, the nuclear modification factor $R_{AA}$ and the elliptic flow $v_2$ of $\text{D}$ mesons, measured at RHIC and the LHC, provide key constraints on this transport coefficient. Numerous theoretical frameworks have been developed to describe these observables consistently~\cite{vanHees:2005wb,vanHees:2007me,Gossiaux:2008jv,Gossiaux:2009mk,Das:2010tj,Uphoff:2012gb,Lang:2012nqy,Alberico:2011zy,Das:2013kea,Das:2015ana,Song:2015sfa,Song:2015ykw,Cao:2013ita,Cao:2016gvr,Katz:2019qwv,Plumari:2019hzp,Singh:2023smw,Krishna:2025bll,Das:2022lqh}.

In addition to the formation of the QGP, it is now well established that extremely intense yet short-lived electromagnetic (EM) fields are generated during the initial stages of high-energy heavy-ion collisions~\cite{Skokov:2009qp,Deng:2012pc,Voronyuk:2011jd,Huang:2015oca}. These fields arise primarily from the motion of spectator protons in non-central collisions and can reach magnitudes of $eB \sim 5$--$50,m_\pi^2$~\cite{Panda:2024fof,Panda:2024ccj,Dash:2023kvr,Panda:2021pvq,Panda:2020zhr,Palni:2024wdy}, several orders of magnitude stronger than the magnetic fields observed on the surfaces of magnetars. As the spectator nuclei recede, the rapidly decaying magnetic field induces strong electric fields, making heavy-ion collisions a unique environment to study QCD matter under extreme electromagnetic conditions~\cite{Panda:2024ccj,Dash:2023kvr,Panda:2021pvq,Panda:2020zhr,Palni:2024wdy,Panda:2025tus}.
Heavy quarks, produced at the earliest stages of the collision and acting as non-equilibrium probes, provide an excellent means to characterize these electromagnetic fields~\cite{Das:2016cwd}. A particularly sensitive observable in this context is the directed flow $v_1$ of heavy mesons~\cite{Das:2016cwd}, which is predicted to be an order of magnitude larger than that of light hadrons. The charge-dependent splitting in the directed flow between charm and anti-charm quarks—studied through $\text{D}^0$ and $\bar{\text{D}}^0$ (or $\text{D}^+$ and $\text{D}^-$) mesons—offers a direct probe of the early-time EM fields~\cite{Das:2016cwd,Chatterjee:2017ahy,Chatterjee:2018lsx,Oliva:2020doe,Oliva:2020mfr,Dubla:2020bdz,Sun:2020wkg,Beraudo:2021ont,Jiang:2022uoe,Sun:2023adv,Das:2025yxy,STAR:2023jdd,Shen:2025unr}.
Experimentally, the STAR and ALICE collaborations have measured the directed flow of $\text{D}$ mesons at RHIC and LHC energies, respectively~\cite{STAR:2019clv,ALICE:2019sgg}. At RHIC, the measured splitting $\Delta v_1 = v_1(\text{D}^0) - v_1(\bar{\text{D}}^0)$ fluctuates around zero within current uncertainties, whereas at the LHC, ALICE reports a positive slope—about three orders of magnitude larger than that observed for light hadrons. Theoretically, most models predict a finite but negative slope at both RHIC and LHC energies. Recent studies suggest that a positive slope in the directed flow splitting of $\text{D}$ mesons at LHC energies can emerge only when the magnetic field dominates over the electric field~\cite{Sun:2020wkg,Jiang:2022uoe}. However, such dominance was introduced in an ad hoc manner to explore possible mechanisms leading to the observed positive slope. These findings emphasize the crucial role of the time evolution of electromagnetic fields, which is strongly governed by the electrical conductivity of the medium—a quantity that remains highly uncertain.

So far, most studies on heavy-quark directed flow have focused exclusively on charm quarks. With recent experimental advances enabling measurements of observables related to $\text{B}$ mesons~\cite{ALICE:2022iba,ALICE:2020hdw}, it becomes timely and compelling to extend these investigations to bottom quarks. Such an analysis offers a unique opportunity to disentangle the effects of both electric charge and mass on the directed flow.
In this work, we study the directed flow $v_1$ and its charge-dependent splitting as functions of rapidity for both charm and bottom quarks. While similar studies exist for the charm sector, we present, for the first time, a comparative analysis of both charm and bottom directed flow, incorporating \textit{all} relevant components of the electromagnetic fields. Moreover, we include the effects of both electrical and chiral conductivities, thereby providing a more comprehensive picture of the underlying dynamics.
The electromagnetic fields are obtained from analytical solutions of Maxwell’s equations with constant electrical and chiral conductivities. Although a fully consistent treatment would ideally require 3+1D magnetohydrodynamic simulations~\cite{Mayer:2024kkv,Mayer:2024dze}, such computations remain technically demanding and computationally intensive. Recent studies have reported charge-splitting effects for light hadrons~\cite{Benoit:2025amn,Panda:2025lmd,Panda:2023akn}; however, our work represents a significant step forward in bridging this gap by extending the analysis to the heavy-quark sector and improving upon existing approaches.

The paper is organized as follows.
The implementation of the complete set of electromagnetic field components, including both electrical and chiral conductivities, is described in Sec.~\eqref{emfields}. The numerical methodology, initialization details, and the Langevin setup are presented in Sec.~\eqref{Numerical setup}. Results for the directed flow of heavy quarks are discussed and analyzed in Sec.~\eqref{Results}, followed by concluding remarks in Sec.~\eqref{conclusions}.

{\bf Notations and Conventions:} 
We use natural units with $\hbar = k_{\rm B} = c = \mu_0 = \epsilon_0 = 1$. Throughout this work, we use the metric with mostly negative signature, i.e., \( g_{\mu \nu} = \mathrm{diag}(+, -, -, -) \).

\section{Electromagnetic field configurations}{\label{emfields}}
Before proceeding to the calculation of the directed flow and the resulting splitting between heavy mesons and their corresponding anti-mesons as a function of rapidity in the presence of electromagnetic fields, we first outline the field configurations employed in this study. The analytical expressions for all field components, derived under the assumption of finite and constant electrical and chiral conductivities, have been extensively discussed in previous works~\cite{Gursoy:2014aka,Siddique:2021smf,Siddique:2022ozg,Panda:2025lmd}. In the present analysis, the electromagnetic fields are evaluated by considering sources propagating along the \(z\)-axis, where the Green’s function method in cylindrical coordinates is employed to obtain the corresponding solutions~\cite{Li:2016tel,Siddique:2021smf,Siddique:2022ozg}.

The physical system under consideration consists of two colliding lead (Pb) nuclei, whose centers are located at $\left(\pm \tfrac{b}{2},\,0,\,0\right)$. 
The nucleus centered at $x_{0} = -\tfrac{b}{2}$ propagates along the negative $z$-direction, while its counterpart at $x_{0} = +\tfrac{b}{2}$ moves along the positive $z$-direction. This choice of convention is a matter of definition and should be applied consistently throughout all simulations. For a given energy, the convention should be fixed and then maintained across different energies to enable meaningful comparison. It is important to note that this convention is opposite to that adopted in many previous works, which followed the conventions of Kharzeev et al.~\cite{Gursoy:2014aka}.
To evaluate the electromagnetic fields at any point on the space--time grid, the spatial distribution of protons within each nucleus is sampled using the Woods--Saxon distribution, $\rho(r) = \rho_0 / [1 + e^{(r - R)/a}]$, where $\rho_0 = 0.16~\mathrm{fm}^{-3}$ is the nuclear density, $R = 6.62~\mathrm{fm}$ is the nuclear radius, and $a = 0.54~\mathrm{fm}$ is the surface diffuseness parameter~\cite{Miller:2007ri}. In this setup, protons are localized in the transverse plane, while their longitudinal motion follows the beam axis ($z$-direction). 
The two nuclei propagate along $\pm z$ with velocity $\pm v t$, where the beam velocity is given by $v = \sqrt{1 - 4 m_p^2 / (\sqrt{s_{\mathrm{NN}}})^2}$, with $m_p$ denoting the proton mass. Transverse coordinates of the protons are generated via Monte Carlo sampling according to the Woods--Saxon distribution. For each event, a total of 164 protons (82 per Pb nucleus) are sampled, with nuclear centers placed at $(\pm b/2,\,0,\,0)$. 
\begin{figure*}[t] 
    \centering
    \begin{subfigure}[b]{0.3\textwidth}
        \includegraphics[width=\textwidth]{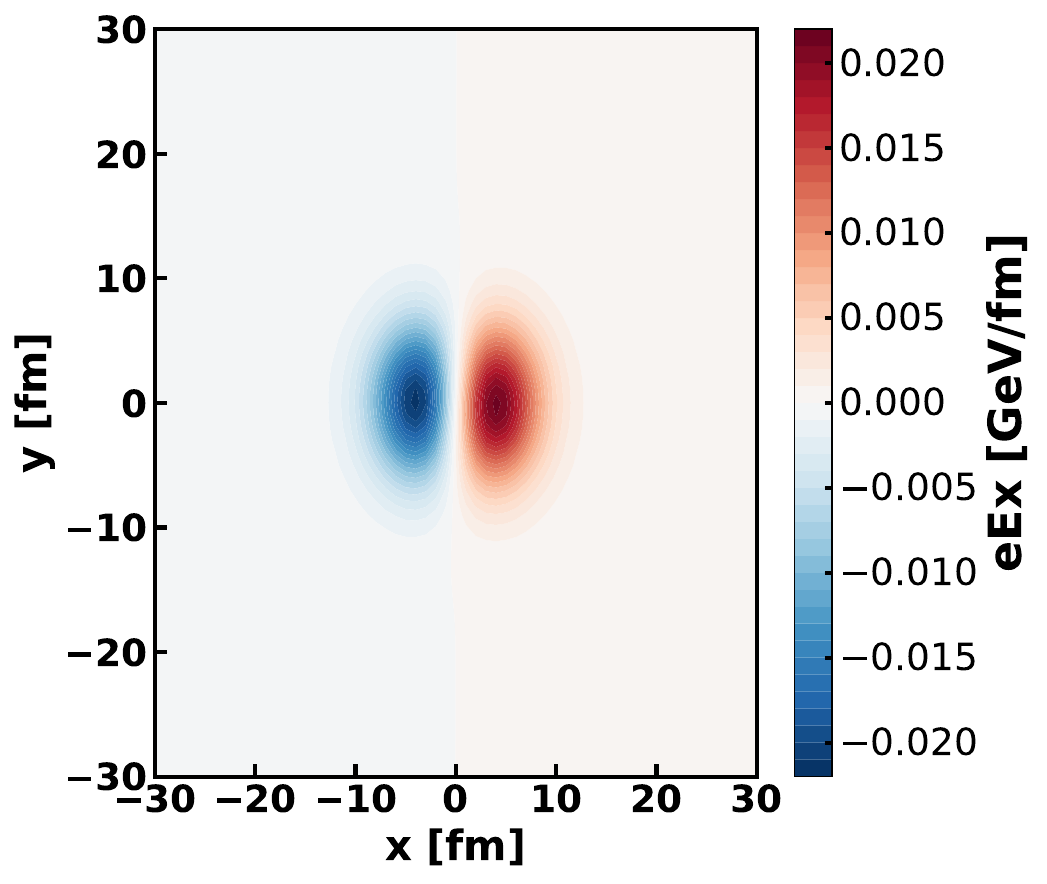}
        \caption{$eE_x$}
    \end{subfigure}\hfill
    \begin{subfigure}[b]{0.3\textwidth}
        \includegraphics[width=\textwidth]{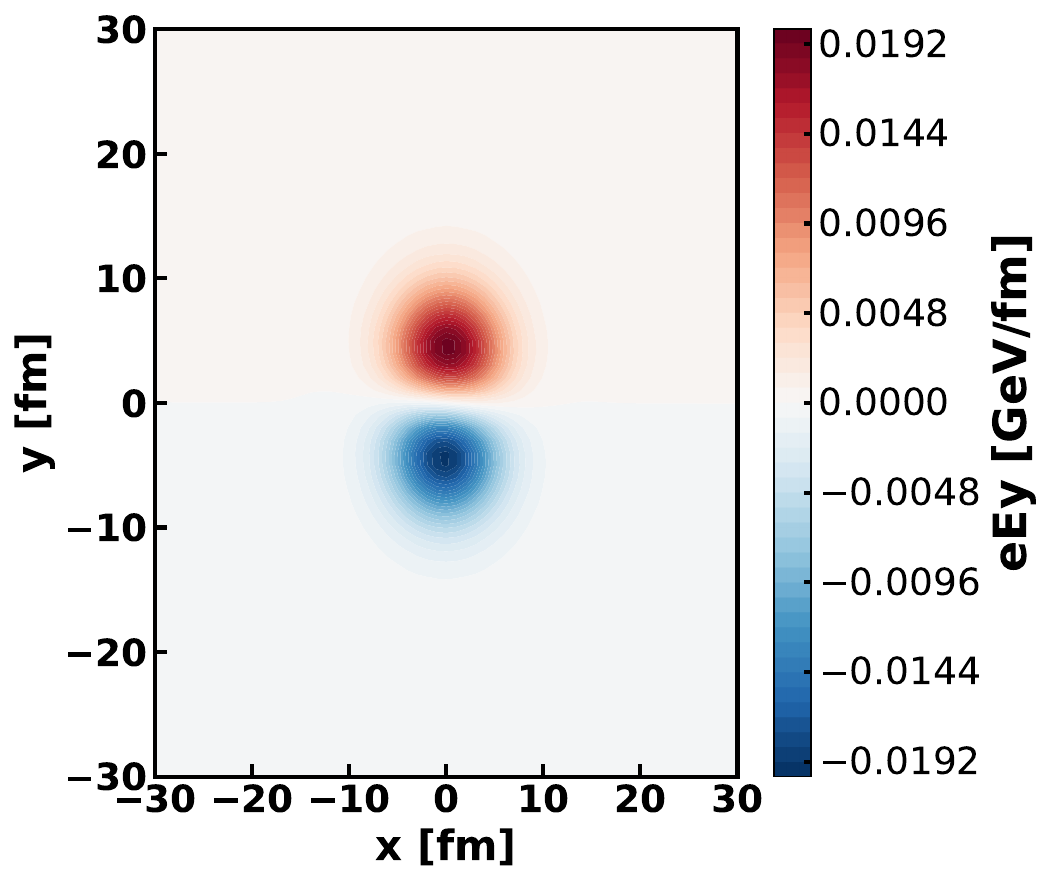}
        \caption{$eE_y$}
    \end{subfigure}\hfill
    \begin{subfigure}[b]{0.3\textwidth}
        \includegraphics[width=\textwidth]{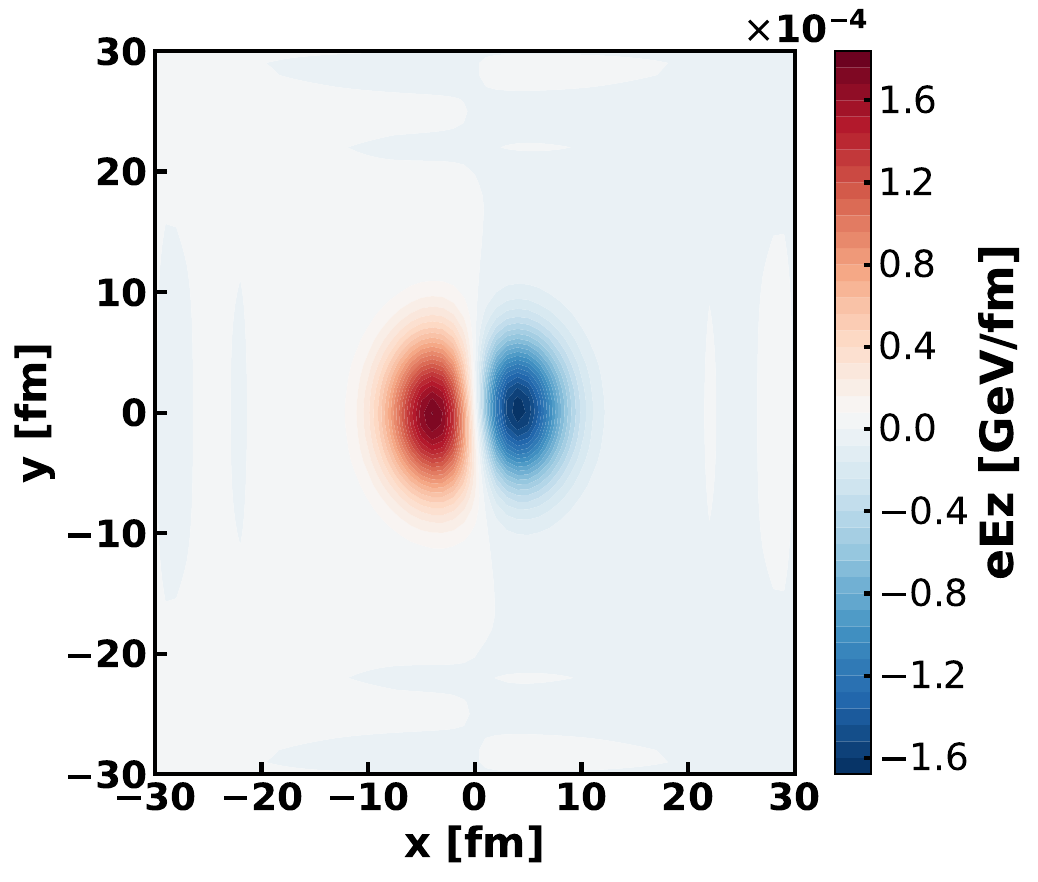}
        \caption{$eE_z$}
    \end{subfigure}

    \vspace{0.4cm}

    \begin{subfigure}[b]{0.3\textwidth}
        \includegraphics[width=\textwidth]{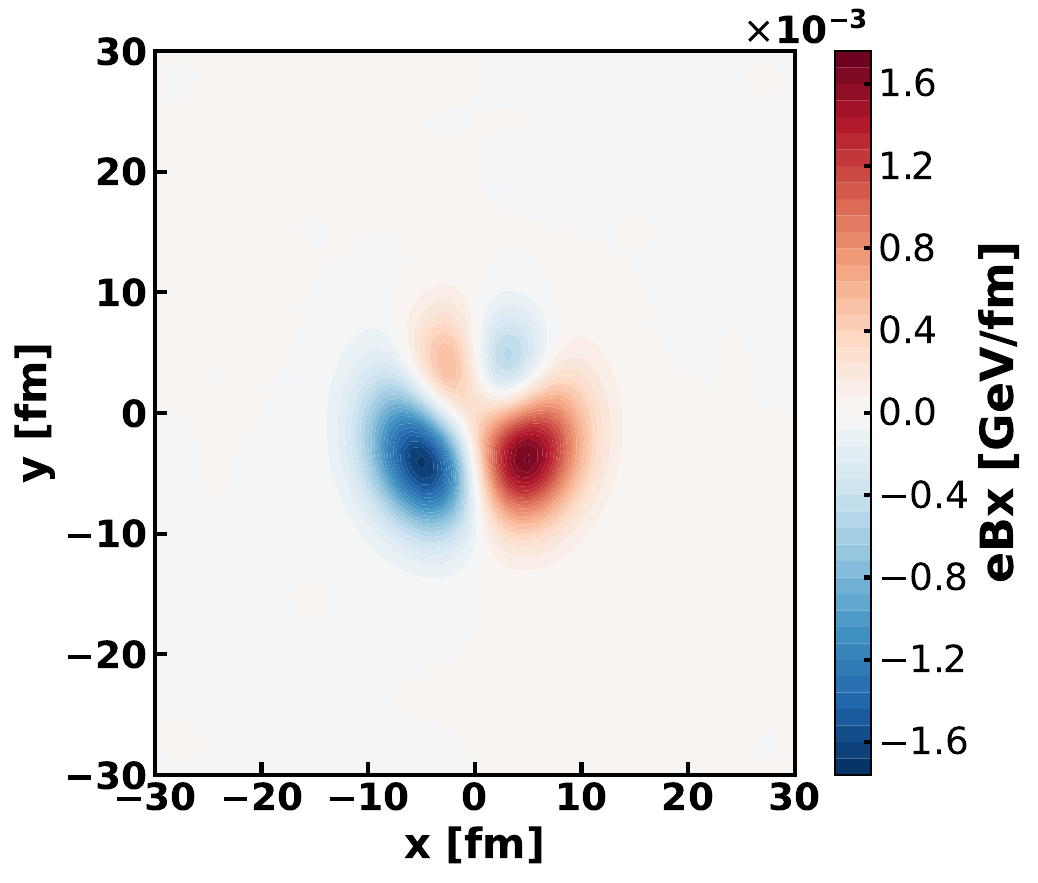}
        \caption{$eB_x$}
    \end{subfigure}\hfill
    \begin{subfigure}[b]{0.3\textwidth}
        \includegraphics[width=\textwidth]{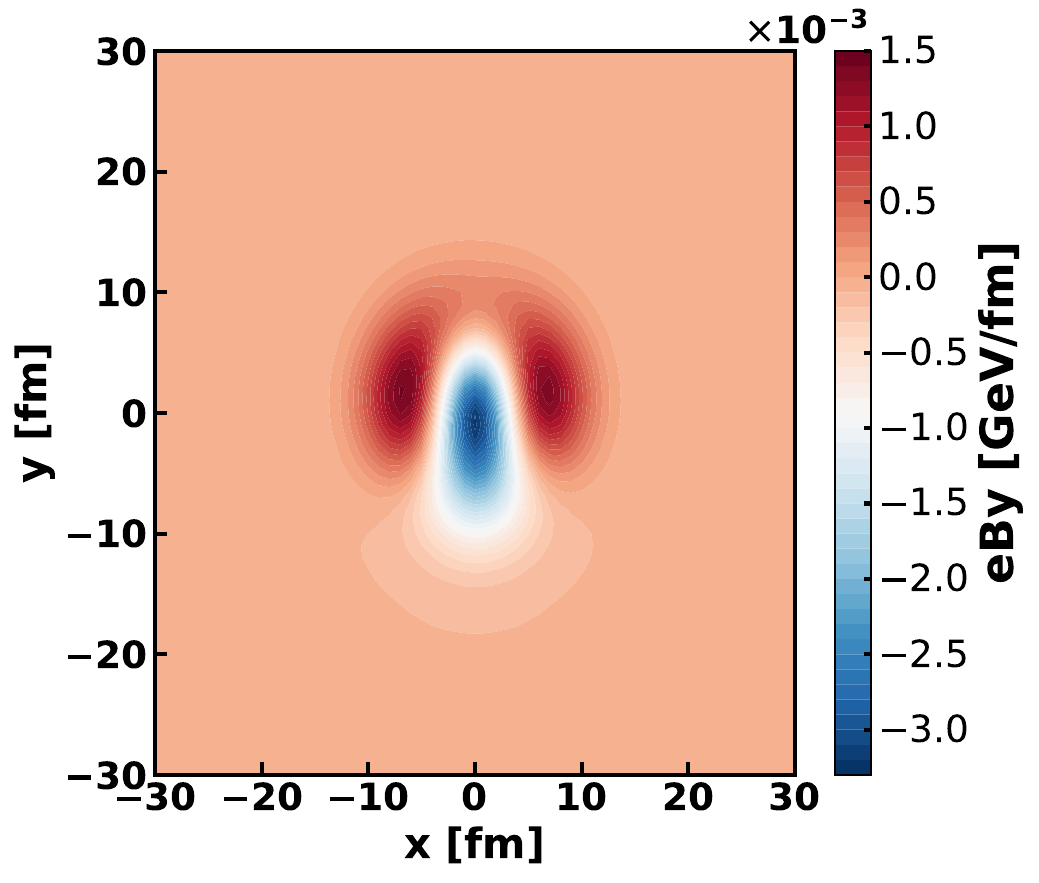}
        \caption{$eB_y$}
    \end{subfigure}\hfill
    \begin{subfigure}[b]{0.3\textwidth}
        \includegraphics[width=\textwidth]{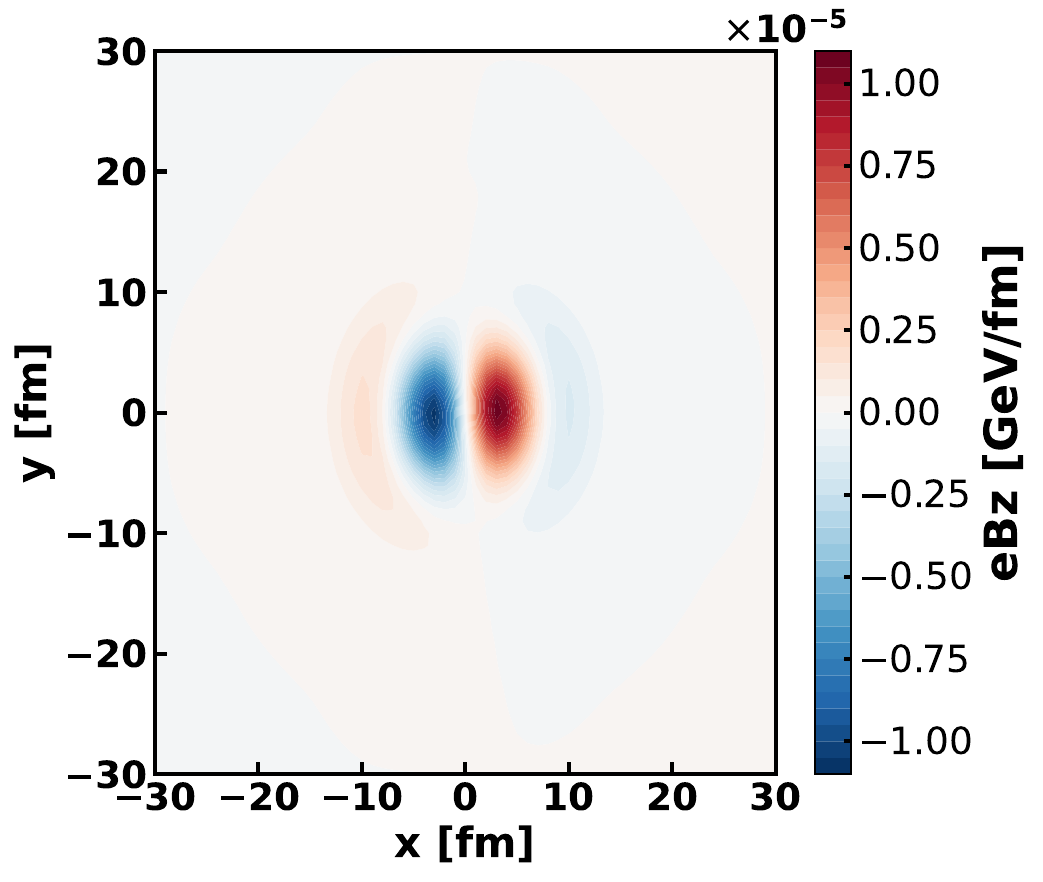}
        \caption{$eB_z$}
    \end{subfigure}

    \caption{Snapshot of all electromagnetic field components at $t = 0.17$~fm for the case including both electrical and chiral conductivity, shown for $b = 9$~fm and $\sqrt{s_{\text{NN}}} = 2.76$~TeV at $\eta = 0$.
}
    \label{fig:Cond1}
\end{figure*}
For each proton pair, the transverse separation is computed as $r_\perp = \sqrt{(x_p - x_T)^2 + (y_p - y_T)^2}$, where $(x_T, y_T)$ and $(x_p, y_p)$ denote the transverse coordinates of the target and projectile protons, respectively. A collision is assumed to occur if $r_\perp \leq r_c$, where the critical radius is defined as $r_c = \sqrt{\sigma_{NN}/\pi}$, with the inelastic nucleon--nucleon cross section $\sigma_{NN}$ obtained at the relevant $\sqrt{s_{\mathrm{NN}}}$ by extrapolating experimental data~\cite{Panda:2024ccj}. Protons identified as participants are assigned their sampled positions and contribute to the medium, thereby generating nonzero electrical and chiral conductivities. This procedure is repeated over 1000 independent events to obtain smooth, event-averaged electromagnetic field configurations in the transverse plane and at the relevant laboratory space--time points.
It is important to note that the presence of finite conductivities—particularly the chiral conductivity—partially breaks the symmetry of the electromagnetic fields. This can be directly inferred from the mathematical structure of the fields discussed in~\cite{Panda:2025lmd}. The origin of this symmetry breaking lies in the additional non-vanishing components introduced by the chiral conductivity $\sigma_{\chi}$, specifically:  
\begin{itemize}  
    \item For the transverse electric field, the inclusion of $\sigma_{\chi}$ generates a nonzero azimuthal component, distorting the otherwise symmetric configuration.  
    \item For the transverse magnetic field, a nonzero radial component emerges in the presence of $\sigma_{\chi}$, leading to a similar symmetry breaking.  
\end{itemize}  
These modifications underscore the role of chiral conductivity in reshaping the electromagnetic field structure and driving deviations from symmetric configurations. A more detailed discussion can be found in~\cite{Li:2016tel,Siddique:2021smf,Siddique:2022ozg}.  
In this work, we adopt $\sigma = 0.023~\mathrm{fm}^{-1}$ and $\sigma_{\chi} = 0.001~\mathrm{fm}^{-1}$ and $0.005~\mathrm{fm}^{-1}$, while spectator protons are assigned zero conductivity. These values are consistent with those used in previous study~\cite{Panda:2025lmd}. The electrical conductivity corresponds to an average value extracted from lattice QCD calculations~\cite{Aarts:2020dda}, whereas the chiral conductivity $\sigma_{\chi}$ has only been indirectly estimated in a few works~\cite{Yamamoto:2011gk,Brandt:2023wgf}. The value of $\sigma_{\chi}$ adopted here lies within the range reported across the highest and lowest temperature regimes. 
This allows for a potential investigation of the sensitivity of both $\sigma$ and $\sigma_{\chi}$ on the resulting electromagnetic fields and, consequently, on bulk observables, which we leave for future work.  
Combining these elements, Figure~\ref{fig:Cond1} shows the transverse profiles of all electromagnetic field components at $t = 0.17~\mathrm{fm}$. 
At earlier times, the fields exhibit a similar qualitative behavior, with larger magnitudes and a more localized spatial distribution. 
As time progresses, the field strength diminishes while its spatial extent broadens, consistent with the expected transverse expansion reported in previous studies~\cite{Li:2016tel,Siddique:2021smf,Siddique:2022ozg}. 
Notably, the symmetry present for $\sigma_{\chi} = 0$ is visibly broken once a finite chiral conductivity is introduced.  

In the following section, we outline the Langevin dynamics of heavy quarks under the influence of these electromagnetic forces.




\section{Langevin equation}{\label{Numerical setup}}

The dynamics of heavy quarks in the QGP are commonly investigated by following their phase-space evolution through a Langevin transport approach.

In the QGP, a heavy quark (HQ) with charge $q$ and momentum $p$ evolves under the influence of an external electromagnetic field according to the relativistic Langevin equation, which can be written as
\begin{eqnarray}
dx_i & = & \frac{p_i}{E} \, dt \ , \\
dp_i & = & -\Gamma(p,T)\, p_i \, dt + C_{ij}(p,T)\,\rho_j \sqrt{dt} + F_{\rm ext} \, dt \ ,
\label{lv1}
\end{eqnarray}
where $dx_i$ and $dp_i$ denote the changes in coordinate and momentum, respectively, over a discrete time step $dt$. Here, $\Gamma(p,T)$ is the drag coefficient, and the covariance matrix is defined as $C_{ij}(p,T) = \sqrt{2B_0 (p,T)}\,P_{ij}^{\perp} + \sqrt{2B_1 (p,T)}\,P_{ij}^{\parallel}$. The transverse and longitudinal projectors are $P_{ij}^{\perp} = \delta_{ij} - p_i p_j/p^2$ and $P_{ij}^{\parallel} = p_i p_j/p^2$, respectively, while $B_0 (p,T)$ and $B_1 (p,T)$ denote the corresponding diffusion coefficients of heavy quarks {depending both on particle momentum and temperature of the medium}. In the limit $p \rightarrow 0$, one has $B_0 = B_1 = D$, leading to $C_{ij} = \sqrt{2D(T)}\,\delta_{ij}$. The stochastic force $\rho$ satisfies $\langle \rho_j \rangle = 0$ and $\langle \rho_i \rho_j \rangle = \delta(t_i - t_j)$.  
The drag and diffusion coefficients, $\Gamma(p,T)$ and $D (p,T)$, encapsulate the microscopic interactions of heavy quarks with the QGP medium. The external Lorentz force acting on a heavy quark is given by $F_{\rm ext} = q\,\mathbf{E} + q\,(\mathbf{p}/E_p) \times \mathbf{B}$, where $E_p$ is the heavy-quark energy. Using the fluctuation-dissipation theorem (FDT), one obtains $D = \Gamma E_p T$, with $T$ denoting the temperature of the thermal bath.  
{In the present work, we employ the pre-point It\^o prescription~\cite{He:2013zua} to solve the Langevin equation, where all momentum-dependent transport coefficients, medium properties, and electromagnetic fields are evaluated at the beginning of each time step using the heavy-quark state at that instant. 
}

{To study the momentum evolution of heavy quarks, we solve the
Langevin equation in an expanding QGP background. A realistic
description of the bulk evolution is required to account for the
expansion, cooling, and build-up of the elliptic flow $v_2(p_T)$.
The QGP medium is modeled using a relativistic partonic transport
approach, initialized with a standard Glauber profile and evolved at
a fixed shear-viscosity-to-entropy-density ratio, $\eta/s=0.16$.
More specifically, the phase-space distributions of light quarks,
antiquarks, and gluons are evolved by solving the relativistic
Boltzmann equation,
\begin{equation}
p^\mu \partial_\mu f_i(x,p)=C_i[f_q,f_{\bar q},f_g],
\qquad i=q,\bar q,g .
\end{equation}
The collision kernel is locally tuned such that the transport
evolution reproduces the prescribed value of $\eta/s$ within the
Chapman--Enskog approximation. This provides a microscopic
realization of a viscous QGP evolution with the chosen value of
$\eta/s$ \cite{Ferini:2008he,Ruggieri:2013bda,Ruggieri:2013ova,Plumari:2012ep,Nugara:2023eku,Sambataro:2025obe,
Sambataro:2025pop,Chen:2026chk}.
At each time step, the system is coarse-grained into spatial cells
to reconstruct the local thermodynamic fields entering the Langevin
coefficients. For each cell, we determine the collective four-velocity
$u^\mu$ and evaluate the energy density $\epsilon$ and particle density
$n$ in the corresponding local rest frame. The local temperature is then
obtained from the equation of state by inverting the relation
$\epsilon(T)/n(T)$. 
For the present study, we investigate the splitting of the directed flow of heavy mesons
in Pb+Pb collisions at $\sqrt{s}=2.76$ ATeV at the LHC. The initial
maximum temperature at the center of the fireball is set to
$T_0=580$ MeV, with the initial time chosen as
$\tau_0\sim 1/T_0=0.15$ fm/$c$.}
{A realistic description of the rapidity-odd directed flow
requires, in general, a three-dimensional initialization of
the bulk medium \cite{Moreland:2014oya}. In particular, a longitudinally tilted
fireball can generate a charge-independent contribution to
the directed flow through the interaction of HQ with the bulk.
Such an analysis is beyond the scope of the present work and will be addressed in future studies.
The main objective of the present work is to isolate the electromagnetic contribution to the charge-dependent $v_1$ splitting and to its dependence on heavy-quark charge and mass.
To leading order, the contribution induced by a tilted but
charge-independent medium is expected to affect heavy mesons
and their antiparticles in the same direction. 
Therefore, while the tilted geometry can modify the rapidity
dependence of the charge-averaged flow, it does not by itself generate a particle--antiparticle
splitting. The correlation between the nucleon position in the nucleus and the stopping of participant matter, together with transverse expansion, can lead to a sign reversal of the directed flow around midrapidity~\cite{Snellings:1999bt}. However, this effect is charge independent and therefore does not contribute to the directed flow splitting.
The latter is primarily driven by the Lorentz force
and is consequently sensitive to the magnitude and time
evolution of the electromagnetic fields
\cite{Das:2016cwd, Chatterjee:2017ahy,Oliva:2020doe,Das:2025yxy}}
The initial momentum distribution of heavy quarks is generated using the Fixed Order plus Next-to-Leading Logarithm (FONLL) calculation~\cite{Cacciari:2012ny}, which reproduces the $\text{D}$-meson spectra observed in proton--proton collisions after fragmentation~\cite{Scardina:2017ipo}. The spatial distribution of charm quarks in coordinate space follows the binary nucleon--nucleon collision profile, $N_{\rm coll}$.  
To describe the interaction of heavy quarks with light quarks and gluons, we employ an extended version of the Quasi‑Particle Model (QPM) \cite{Plumari:2011mk,Das:2015ana,Scardina:2017ipo,Plumari:2019hzp,Sambataro:2023tlv}, referred to as $QPM_p$. This approach incorporates temperature and momentum‑dependent parton masses, consistent with QCD asymptotic freedom, and has been shown to reproduce the lattice‑QCD Equation of State as well as the susceptibilities of light, strange, and charm quarks \cite{Berrehrah:2015vhe,Berrehrah:2016vzw,Sambataro:2024mkr}. In this framework, the drag coefficient at finite momentum, $\Gamma(p,T)$, is computed from the scattering matrices $|\mathcal{M}_{HQ+g(q)\rightarrow HQ+g(q)}|$ evaluated at tree level, while the effective vertex coupling $g(T)$ is extracted from a fit to the lattice‑QCD energy density $\epsilon$. The resulting coupling is significantly larger than that obtained in perturbative QCD, particularly as $T\rightarrow T_c$.  The spatial diffusion coefficient $D_s(p \rightarrow 0)$ in $QPM_p$ is close to the new lattice-QCD data near $T_c$ \cite{Altenkort:2023oms,Altenkort:2023eav,HotQCD:2025fbd}, especially for bottom quark \cite{Sambataro:2024mkr,Sambataro:2025obe}.  This model successfully describes the nuclear modification factor $R_{AA}$, the elliptic flow $v_2$ and the triangular flow $v_3$ of D-mesons (see Ref.\cite{Sambataro:2025obe} for further discussion). Notice that for the interaction of bottom quarks with the thermal medium, we employ the same $QPM_p$ with identical parameters with respect to charm quarks, replacing only the charm quark mass with that of the bottom quark.
At the end of the QGP phase, when the bulk temperature drops below the quark--hadron transition temperature, charm and bottom quarks are converted into $\text{D}$-mesons and $\text{B}$-mesons using the Peterson fragmentation function~\cite{Scardina:2017ipo}.  
Finally, the directed flow of heavy quarks is computed as $v_1 = \langle \cos(\phi) \rangle = \langle p_x / p_T \rangle$, where $\phi$ denotes the azimuthal angle.


\section{Results}{\label{Results}} 

\begin{figure}[htbp]
    \centering
    \begin{minipage}[b]{0.5\textwidth}
        \centering
        \includegraphics[width=\textwidth]{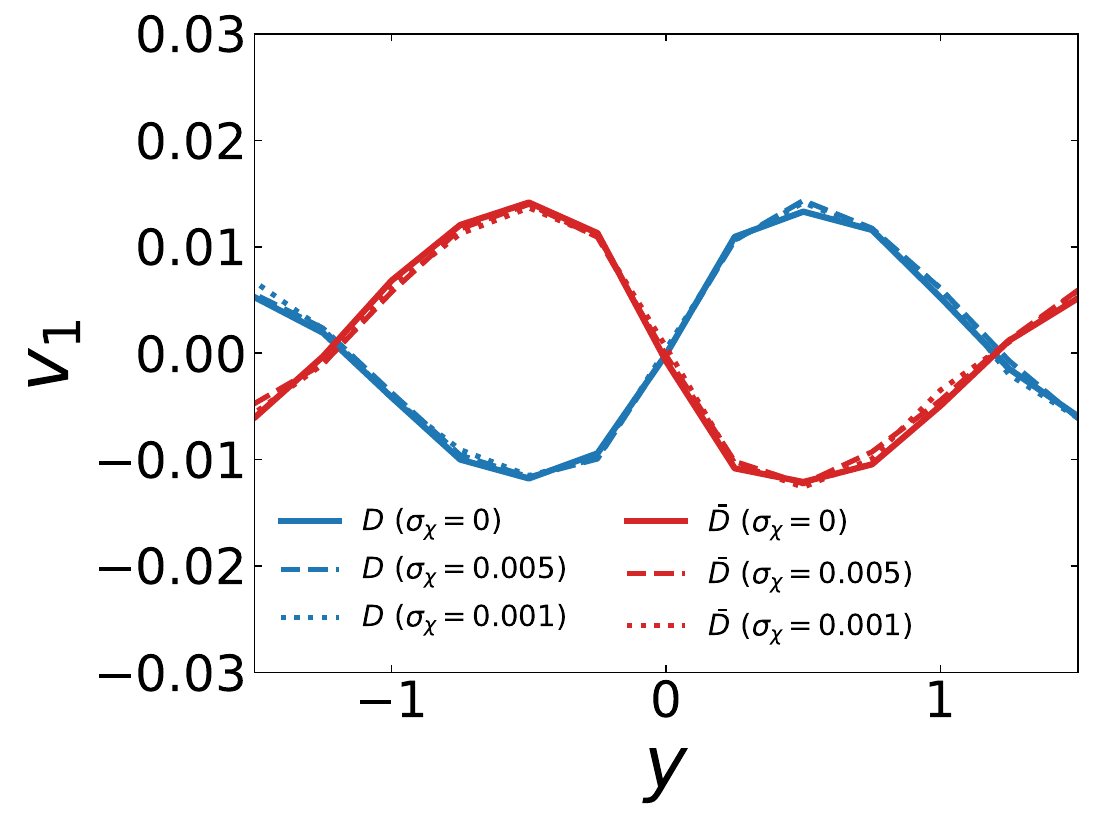}
        \label{fig:one}
    \end{minipage}
    \caption{Rapidity dependence of directed flow of $\text{D}$ and $\bar{\text{D}}$ mesons with (dashed) and without (solid) chiral conductivity.}
    \label{dmeson}
\end{figure}

\begin{figure}[htbp]
    \centering
    \includegraphics[width=0.5\textwidth]{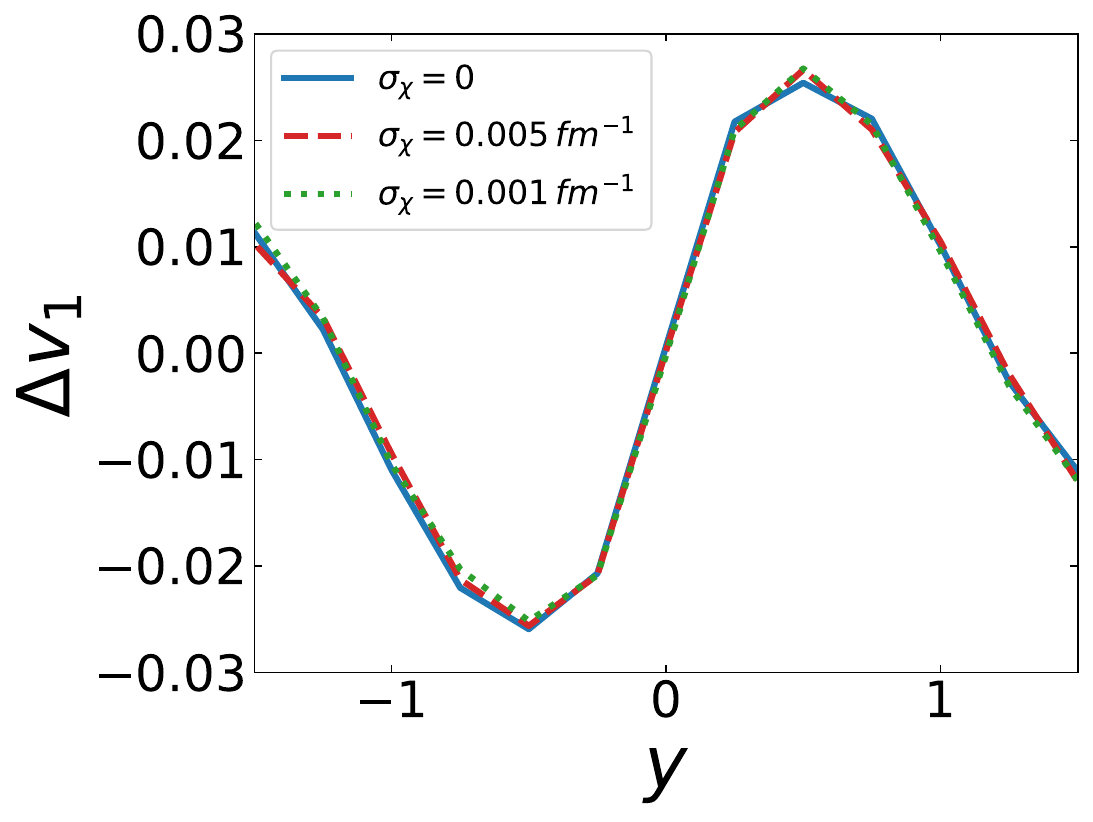}
     \caption{Rapidity dependence of $\Delta v_1$ ($v_1(\text{D})-v_1(\bar{\text{D}})$) for $\text{D}$ mesons with (dashed) and without (solid) the chiral conductivity.}
    \label{dsplit}
\end{figure}

With all essential components established, we solve the Langevin equation in the presence of electromagnetic fields, incorporating the effects of chiral conductivity, to compute the directed flow $v_1$ of heavy mesons. For clarity and conciseness, we do not detail the standard procedures for evaluating momentum and coordinates in the proper frame or the subsequent Lorentz transformations required to obtain observables in the laboratory frame, as these methods are well documented in the literature. Within this framework, we determine the directed flow $v_1$ of both $\text{D}$ and $\text{B}$ mesons, enabling a systematic investigation of the charge and mass dependence of heavy‑meson directed flow.
Figure~\eqref{dmeson} shows the directed flow of $\text{D}$ $[c\bar{q}]$ and $\bar{\text{D}}$ $[q\bar{c}]$ mesons obtained after the fragmentation of charm quarks at \(\sqrt{s_{\text{NN}}} = 2.76\)~TeV for collisions with an impact parameter of $b = 9$~fm. The simulations were performed using an electrical conductivity of $\sigma = 0.023~\text{fm}^{-1}$ and a chiral conductivity of $\sigma_{\chi} = 0.001~\text{fm}^{-1}$ and $0.005~\text{fm}^{-1}$. {It should also be noted that the electromagnetic fields used in the simulations are event-averaged fields entering our calculations, rather than event-by-event electromagnetic fields produced in each realization of the Langevin evolution.} To study the rapidity dependence, we integrate over the transverse momentum $p_T$ of the heavy mesons. In the figure~\eqref{dmeson}, the blue curves correspond to the $\text{D}$ mesons, while the red curves represent the $\bar{\text{D}}$ mesons $[\bar{c}q]$ obtained after fragmentation of the subsequent quarks. Both sets include the effects of electrical conductivity: the solid lines correspond to calculations without chiral conductivity, whereas the dashed lines include it. The $\text{D}$ mesons exhibit a negative directed flow at negative rapidity, which changes sign at positive rapidity. For the $\bar{\text{D}}$ mesons, the sign of the directed flow is reversed, primarily due to their opposite electric charge and the corresponding coupling to the electromagnetic fields. Here, we note that the sign of the directed flow as a function of rapidity obtained in this calculation is opposite to that reported in~\cite{Das:2016cwd, Oliva:2020doe}. This difference originates from the convention adopted for the electromagnetic field configuration in the present study. A change in the chosen field convention naturally leads to an opposite sign in the directed flow as a function of rapidity. It is therefore essential that the convention be fixed consistently, allowing meaningful exploration and comparison across different collision energies and impact parameters.
Figure~\eqref{dsplit} shows the variation of the directed flow splitting, $\Delta v_1 = v_1(\text{D}) - v_1(\bar{\text{D}})$, as a function of rapidity with and without chiral conductivity. The inclusion of chiral conductivity has only a marginal effect. These small differences arise because chiral conductivity contributes as a next-to-leading-order (NLO) correction to the electromagnetic fields, while the dominant effect continues to come from the electrical conductivity, which plays a crucial role in prolonging the lifetime of the fields.
The splitting induced by electromagnetic fields at the LHC remains finite, with the electric field dominating at lower rapidities. This behavior 
determines the sign of the splitting. At $\sqrt{s_{\text{NN}}} = 2.76$~TeV, no experimental measurements are currently available; existing data have been reported only at the higher energy of $\sqrt{s_{\text{NN}}} = 5.02$~TeV~\cite{ALICE:2019sgg}. Consequently, we do not perform a direct comparison with experimental results. Nevertheless, the magnitude of the predicted splitting is found to be comparable within the experimental uncertainties. We emphasize that a combined investigation of the directed flow of both charm and bottom quarks could provide valuable insight into the electromagnetic origin of directed flow in the heavy quark sector.

\begin{figure}[htbp]
    \centering
        \centering
        \includegraphics[width=0.5\textwidth]{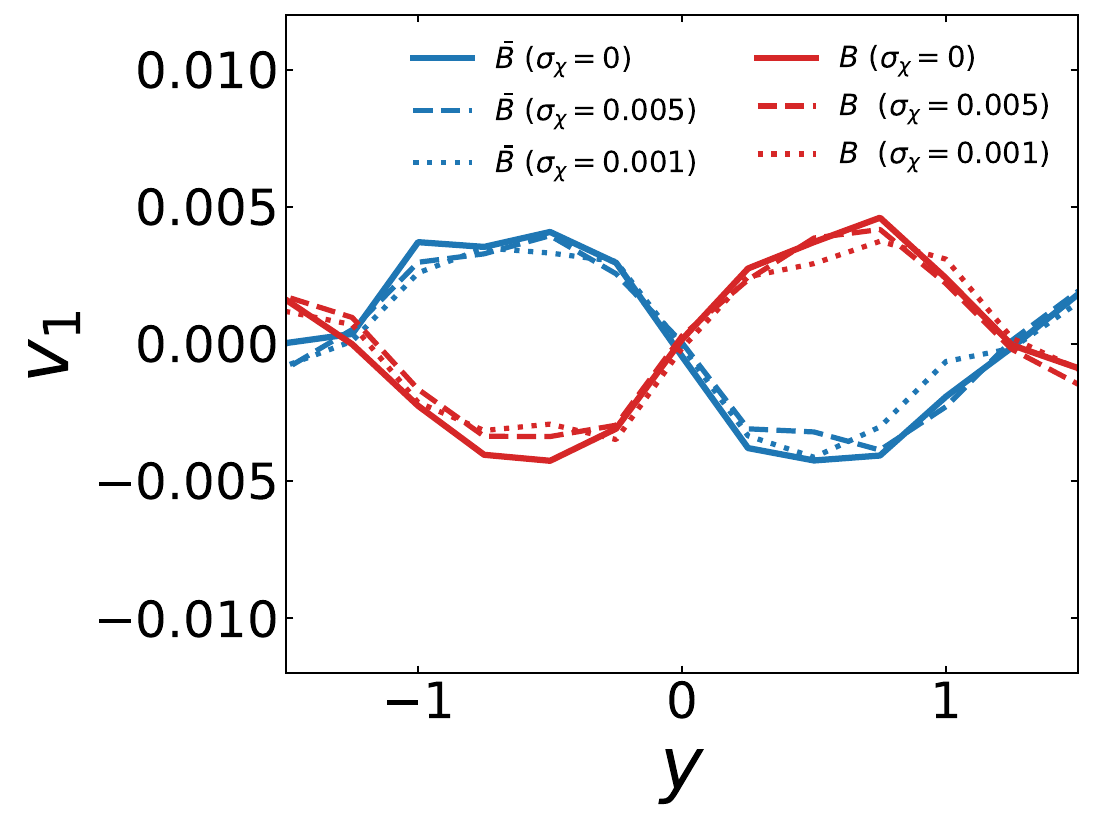}
        \label{fig:one}
    \caption{Rapidity dependence of directed flow of $\text{B}$ and $\bar{\text{B}}$ mesons with (dashed) and without (solid) chiral conductivity.}
    \label{bmeson}
\end{figure}

\begin{figure}
    \centering
    \includegraphics[width=0.5\textwidth]{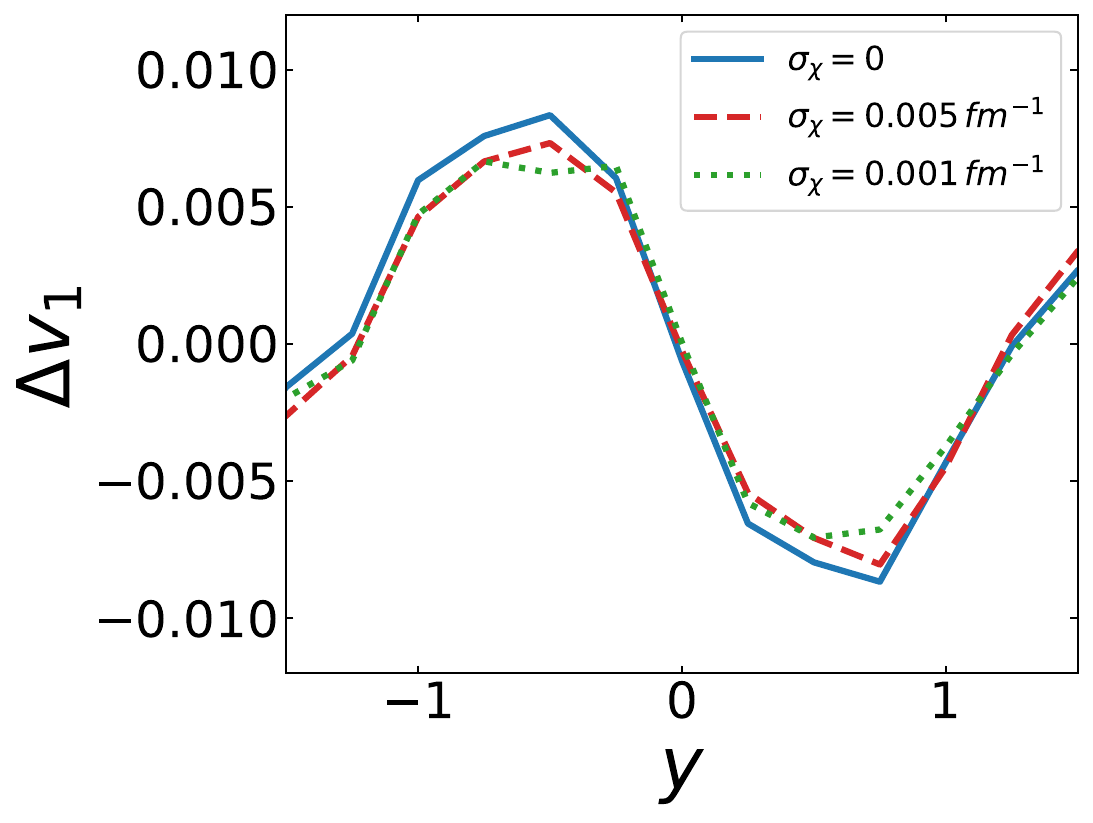}
    \caption{Rapidity dependence of $\Delta v_1$ ($v_1(\bar{\text{B}}) $- $v_1(\text{B}$)) for $\text{B}$ mesons with (dashed) and without (solid lines) the chiral conductivity.}
    \label{bsplit}
\end{figure}

Next, we focus on the ${\text{B}}$ mesons—representing the first calculation for mesons containing a bottom quark that treats both $\text{D}$ and $\text{B}$ mesons within a unified framework. Figure~\eqref{bmeson} presents the rapidity dependence of the directed flow for $\bar{\text{B}}$ and $\text{B}$ mesons, with and without the effects of chiral conductivity. The blue curve corresponds to the $\bar{\text{B}}$ mesons ($[{\text{b}}\bar{\text{q}}]$), while the red curve represents the $\text{B}$ mesons ($[\bar{\text{b}}{\text{q}}]$). Interestingly, the trend observed for $\bar{\text{B}}$ mesons, originating from the fragmentation of bottom quarks, is opposite to that seen for $\text{D}$ mesons produced from charm quarks. This inversion arises primarily from the difference in quarks' electric charges ($+2/3$ for charm and $-1/3$ for bottom) when transitioning from the charm to the bottom sector, highlighting the crucial role of quark charge in interactions with the electromagnetic field. In the absence of electromagnetic fields, no reversal in sign would be expected between the $\text{D}$ and $\bar{\text{B}}$ mesons.  
Moreover, the overall magnitude of the directed flow for $\bar{\text{B}}$ mesons is smaller than that for $\text{D}$ mesons, reflecting the fact that the bottom quark carries an electric charge smaller by a factor of two compared to the charm quark. The quark mass also plays an important role: while a heavier quark has a longer relaxation time, which could enhance $v_1$, its larger mass simultaneously suppresses the contribution from the magnetic field. Thus, both quark charge and mass significantly influence the overall behavior. These results indicate that charm quarks act as more effective probes of the directed flow ($v_1$) compared to bottom quarks, primarily due to their larger electric charge. However, we also observe that the impact of chiral conductivity is more pronounced for $\bar{\text{B}}$ mesons than for $\text{D}$ mesons, which can be attributed to the longer thermalization time of $\bar{\text{B}}$ mesons. 
In Figure~\eqref{bsplit}, we present the variation of the directed flow splitting, defined as $\Delta v_1 = v_1(\bar{\text{B}}) - v_1(\text{B})$, as a function of rapidity, both with and without the inclusion of chiral conductivity. A finite splitting is observed for the bottom quark; however, its magnitude is smaller than that of the charm quark, primarily due to the smaller electric charge of the bottom quark. Another noteworthy observation in the directed flow results (Figure~\eqref{dmeson} and Figure~\eqref{bmeson}) is the emergence of a secondary crossing at higher rapidity, in addition to the crossing observed at midrapidity. This feature is also evident in the splitting plots of directed flow shown in Figure~\eqref{dsplit} and Figure~\eqref{bsplit}. The occurrence of this secondary crossing is a natural consequence of the competition between the dominant electromagnetic field components ($B_y$ and $E_x$). The precise rapidity at which the crossing appears is model-dependent, being sensitive to the spatial distribution of the electromagnetic fields, which in the present case is determined by the Woods–Saxon potential. In our simulations, the electric field dominates at lower rapidities, while the magnetic field gradually becomes stronger at higher rapidities. This transition provides the physical basis for the secondary crossing observed in the directed flow.



\section{Summary and conclusions}{\label{conclusions}}
This study presents a comprehensive framework for investigating the directed flow \(v_1\) and charge-dependent splitting \(\Delta v_1\) of heavy mesons in relativistic heavy-ion collisions, incorporating the effects of electromagnetic fields with chiral conductivity. The interactions of heavy quarks with light quarks and gluons in the QGP are modeled using an extended version of the Quasi-Particle Model, denoted as $QPM_p$, which incorporates non-perturbative effects through temperature- and momentum-dependent quasi-particle masses and couplings. We focus specifically on $\text{D}$ and $\text{B}$ mesons, treated on an equal footing for the first time.
We begin by analyzing $\text{D}$ mesons at a collision energy of \(\sqrt{s_{\text{NN}}} = 2.76\)~TeV at impact parameter of 9 fm. The directed flow exhibits a negative value at negative rapidity, which changes sign at positive rapidity. The introduction of chiral conductivity however produces only a marginal effect, indicating that it acts as a next-to-leading-order (NLO) correction to the dominant influence of electrical conductivity, which primarily governs the lifetime and evolution of the electromagnetic fields in the system.
Next, we extend the analysis to $\text{B}$ mesons, marking the first calculation where both $\text{D}$ and $\text{B}$ mesons directed flow and their splitting in presence of electromagnetic fields are shown. In contrast to $\text{D}$ mesons, $\bar{\text{B}}$ mesons exhibit an opposite trend in their directed flow, primarily due to the difference in electric charge between charm and bottom quarks. The magnitude of the directed flow for $\bar{\text{B}}$ mesons is suppressed relative to $\text{D}$ mesons, reflecting the smaller electric charge of the bottom quark. Interestingly, the impact of chiral conductivity is more pronounced for $\bar{\text{B}}$ mesons than for $\text{D}$ mesons, which can be attributed to the longer thermalization time of bottom quarks. We also observe a secondary crossing in the rapidity dependence of the directed flow of heavy quarks. This feature arises primarily from the interplay between the rapidity variations of the $E_x$ and $B_y$ field components.


In conclusion, this study advances the understanding of heavy meson dynamics in relativistic heavy-ion collisions by presenting a unified framework for $\text{D}$ and $\text{B}$ mesons that incorporates $QPM_p$-based heavy-quark interactions together with electric and chiral conductivity effects. While electrical conductivity remains the dominant factor shaping the directed flow of mesons, the inclusion of chiral conductivity introduces more subtle modifications to the system, particularly in the case of $\text{B}$ mesons. The results highlight the contrasting behaviors of $\text{D}$ and $\bar{\text{B}}$ mesons in the presence of electromagnetic fields. A simultaneous experimental measurement of $v_1$ for heavy mesons containing both charm and bottom quarks would however provide valuable insights into the electromagnetic-field origin of $v_1$ in the heavy-quark sector. 

We emphasize that the present study provides a controlled baseline framework for investigating chiral electromagnetic effects in heavy-flavor observables. The treatment of chiral conductivity is based on the hierarchy $\sigma_{\chi}\ll\sigma_{\rm el}$, where $\sigma_{\chi}$ is included as a next-to-leading-order correction to the electrical conductivity, with both conductivities treated as constant parameters throughout the evolution. Consequently, the electromagnetic fields are obtained from an analytical solution, and the resulting chiral effects should be interpreted within this perturbative framework. In future work, it would be worthwhile to employ a fully dynamical MHD or anomalous MHD framework, in which the electromagnetic fields and chiral transport evolve self-consistently without these approximations. Such an approach would naturally incorporate effects such as event-by-event fluctuations and time-dependent chiral conductivity, providing a more complete and realistic description of the underlying dynamics.

Nevertheless, the present framework provides valuable insight into the interplay between heavy-quark flavor, mass, electromagnetic fields, and chiral transport effects, and establishes a foundation for more comprehensive studies of heavy-meson directed flow in relativistic heavy-ion collisions.

\section{Acknowledgment}
Ankit Kumar Panda is supported by Central China Normal University and by NSFC Grants No.12435009.
Pooja acknowledges the support of the Research Council of Finland, the Centre of Excellence in Quark Matter (project 346324 and 364191), and by the European Research Council (ERC, grant agreement No. ERC-2018-ADG835105 YoctoLHC). SKD acknowledges the support from Anusandhan National Research Foundation (ANRF), India, under grant No.:ANRF/ARG/2025/002424/PS

\bibliographystyle{unsrt}
\bibliography{reference}

\end{document}